\documentclass[journal,comsoc]{IEEEtran}
\usepackage{amsmath}
\usepackage[cmintegrals]{newtxmath}
\usepackage{placeins} 
\usepackage{underscore} 
\usepackage{lipsum}
\usepackage[font=itshape]{quoting} 
\usepackage{graphicx} 
\usepackage{subcaption}
\usepackage{textcomp} 
\usepackage{amsmath} 
\usepackage{pbox} 
\usepackage{hyperref} 
\hypersetup{draft}
\usepackage{xfp} 
\usepackage{txfonts}
\usepackage{colortbl} 
\usepackage{array, makecell}
\usepackage{booktabs}
\usepackage{makecell, booktabs}
\usepackage{tabularx} 
\usepackage[para,online,flushleft]{threeparttable} 
\usepackage{multicol} 
\usepackage{multirow} 

\usepackage{amssymb}
\usepackage{pifont}

\usepackage{tabularx}
\usepackage{threeparttable, tablefootnote}
\usepackage{wrapfig} 
\usepackage{adjustbox} 
\usepackage{paralist} 
\usepackage{msc} 
\newcommand{\quotes}[1]{``#1''} 
\usepackage{mathptmx} 

\usepackage[inline]{enumitem}

\setlist[enumerate, 2]{
	label=\theenumi.\arabic*),
	ref=\theenumi.\arabic*,
	leftmargin=*}

\usepackage{sistyle} 
\SIthousandsep{,} 

\usepackage{colortbl} 

\definecolor{lightYellow}{cmyk}{0,0.04,4.4,0}
\definecolor{lightPurple}{cmyk}{0,0.27,0,0.07}
\definecolor{lightGray}{cmyk}{0,0,0,0.8}
\usepackage{xcolor} 
\usepackage{authblk}
\usepackage{flushend} 

\begin{document}
\title{Exploring HTTPS Security Inconsistencies:\\A Cross-Regional Perspective}

\author[1,2]{Eman Salem Alashwali}
\author[3]{Pawel Szalachowski}
\author[1]{Andrew Martin}
\affil[1]{University of Oxford, UK \authorcr Email: {\tt \{eman.alashwali|andrew.martin\}@cs.ox.ac.uk}\vspace{1ex}}
\affil[2]{King Abdulaziz University, Saudi Arabia \authorcr Email: {\tt ealashwali@kau.edu.sa}\vspace{1ex}}
\affil[3]{SUTD, Singapore \authorcr Email: {\tt pawel@sutd.edu.sg}} 
\maketitle

\begin{abstract}
If two or more identical HTTPS clients, located at different geographic locations (regions), make an HTTPS request to the same domain (e.g. \texttt{example.com}), on the same day, will they receive the same HTTPS security guarantees in response? Our results give evidence that this is not always the case. We conduct scans for the top \num{250000} most visited domains on the Internet, from clients located at five different regions: Australia, Brazil, India, the UK, and the US. Our scans gather data from both application (URLs and HTTP headers) and transport (servers' selected TLS version, ciphersuite, and certificate) layers. Overall, we find that HTTPS inconsistencies at the application layer are higher than those at the transport layer. We also find that HTTPS security inconsistencies are strongly related to URLs and IPs diversity among regions, and to a lesser extent to the presence of redirections. Further manual inspection shows that there are several reasons behind URLs diversity among regions such as downgrading to the plain-HTTP protocol, using different subdomains, different TLDs, or different home page documents. Furthermore, we find that downgrading to plain-HTTP is related to websites' regional blocking. We also provide attack scenarios that show how an attacker can benefit from HTTPS security inconsistencies, and introduce a new attack scenario which we call the \quotes{region confusion} attack. Finally, based on our analysis and observations, we provide discussion, which include some recommendations such as the need for testing tools for domain administrators and users that help to mitigate and detect regional domains' inconsistencies, standardising regional domains format with the same-origin policy (of domains) in mind, standardising secure URL redirections, and avoid redirections whenever possible.
\end{abstract}

\section{Introduction} \label{sec:intro}
\subsection{Motivation}
The Transport Layer Security (TLS) protocol is one of the most important and widely used protocols to date. It provides integrity, secrecy, and authentication for application layer protocols such as the HyperText Transfer Protocol (HTTP), which carries sensitive data such as those transferred in e-banking, e-government, and e-commerce applications. Since its early versions, TLS, especially in the context of HTTPS, has been receiving scrutiny from the security research community. This, in turn, revealed a large number of vulnerabilities. As a result, new TLS versions, ciphersuites, HTTP security headers, and guidelines, have been introduced to achieve the desired security of HTTPS. \par        

With the existence of multiple TLS versions, ciphersuites, and HTTP security headers, it is important to continuously assess the real world deployment of HTTPS security configurations. Several studies have assessed some aspects of HTTPS deployment in the real world, such as~\cite{durumeric13}\cite{felt17}\cite{amann17}. However, the quest for assessing the consistency of servers' HTTPS security guarantees in response to requests from different geographic regions for the same domain remains largely overlooked. \par 

\subsection{Research Question}
In this paper, we aim to answer the following question: \textit{If two or more identical HTTPS clients, located at different regions, make an HTTPS request to the same domain, on the same day, will they receive the same HTTPS security guarantees in response?} \par 

In particular, we investigate the security guarantees from three vectors: \begin{inparaenum}[1)] \item URLs security, \item Security headers, and \item TLS security, \end{inparaenum}with respect to certain security properties, chosen due to their perceived importance. \autoref{tab:vectors} list the vectors and properties we examine.

\begin{table}[!tp]
\centering
\caption{The three vectors along with the security properties that we examine in our research question.}	
\label{tab:vectors} 
\begin{tabular}{>{}l>{}l}
	\toprule 
	Vector & Property \\
	\midrule 
	
	\multirow{4}{*}{URL Security} & HTTPS in final URLs \\
								  & HTTPS in intermediate URLs \\
								  & Compatible domains in final URLs \\
								  & Compatible domains in intermediate URLs \\
								  
	\hline 
	\multirow{3}{*}{Security Headers} & HSTS headers\\
										& CSP headers\\
										& The Secure attribute in Set-Cookie headers \\
	\hline 
	\multirow{6}{*}{TLS Security} & Protocol version TLS~1.3\\
								  & Protocol version TLS~1.2 or higher \\
								  & The Forward Secrecy (FS) property \\
								  & The FS and Authenticated Encryption (AE) properties \\
								  & Non-expired certificates \\
								  & Valid hostnames in certificates \\
	\bottomrule		
\end{tabular}	
\end{table}
	
\subsection{Contributions}
The contributions of this paper are as follows: first, we provide the first assessment for the inconsistencies of servers' HTTPS security guarantees in response to requests for the same domain from different regions. That is, we quantify servers that exhibit weakness indicators in their responses to \textit{\textbf{some but not all}} regions. Second, we study the relationship between HTTPS security inconsistencies and URLs diversity, IPs diversity, and the presence of redirections. Third, we introduce a novel attack scenario that we call the \quotes{region confusion} attack. In this attack, the attacker exploits HTTPS security inconsistencies in servers' responses, and redirects a client's request, e.g. via DNS spoofing, from a region with strong HTTPS security, to another region with weak HTTPS security, to exploit the weaker region's weaknesses. 

\subsection{Organisation}
The rest of the paper is organised as follows: In \autoref{sec:background}, we provide a brief background and the threat model. In \autoref{sec:empirical}, we describe our experiment. In \autoref{sec:results}, we summarise our results. In \autoref{sec:scenario}, we provide two attack scenarios. In \autoref{sec:discussion}, we provide recommendations. In \autoref{sec:related}, we summarise related work. We list some limitations in \autoref{sec:limitations}. Finally, in \autoref{sec:conclusion}, we conclude our research.

\section{Background and Preliminary} \label{sec:background}
\subsection{TLS, HTTP, and HTTPS}
The TLS protocol~\cite{rescorla08tls12}\cite{rescorla18tls13}, formerly known as SSL, is one of the most widely used protocols to date. It has been in use since 1996. The latest and sixth version of TLS is known as TLS~1.3 \cite{rescorla18tls13}, which was standardised in August 2018. TLS operates below application layer protocols such as the HTTP protocol, to provide them with a secure channel. HTTP on its own is insecure. Indeed, websites running plain-HTTP are prone to eavesdropping, manipulation, and impersonation attacks. TLS consists of multiple sub-protocols including the handshake protocol, which is the most security-critical part of the TLS protocol. In the handshake protocol, both communicating parties negotiate the protocol version and the ciphersuite that will be used in subsequent messages of the protocol, authenticate each other\footnote{In our paper, we only consider unilateral server authentication, where only the client authenticates the server.}, and exchange the session keys. The ciphersuite is an identifier, represented by a string or a
hexadecimal value, which defines the cryptographic algorithms (e.g. symmetric encryption) and their parameters (e.g. key length) that will be used in subsequent messages of the protocol. The negotiation of the protocol version and ciphersuite is intrinsic. It defines the security guarantees that the protocol can provide in a particular session. Some ciphersuites provide stronger security guarantees than others. For example, some ciphersuites provide the Forward Secrecy (FS) property (FS-ciphersuites) which provides resilience against mass surveillance, or the Authenticated Encryption (AE) property (AE-ciphersuites) which protects against attacks over the MAC-then-encrypt schemes, or both FS and AE (FS+AE-ciphersuites), while some other ciphersuites neither provide FS nor AE (non-FS+non-AE-ciphersuites). The same applies to TLS versions, where version TLS~1.3 provides stronger guarantees than older versions, which have known design weaknesses. 

\subsection{HTTP Redirection}
HTTP redirection, also known as URL redirection~\cite{http14}, is a technique that allows a resource (e.g. a website) to be reachable from multiple URL addresses. URL redirection  also includes domain redirection, where a website is reachable from one or more domain names. For example, when a website can be reached through either its plain-domain (e.g. \texttt{example.com}) or its equivalent www-domain (e.g. \texttt{www.example.com}), it is because requests to the plain-domain are redirected to its equivalent www-domain, or vice versa. Similarly, domain redirection is used when requests to a domain with a generic Top Level Domain (gTLD) (\texttt{example.com}) are redirected to a regional domain with a country code TLD (ccTLD) (e.g. \texttt{example.co.uk}), or to a regional subdomain (e.g. \texttt{uk.example.com}). HTTP redirection can be achieved through various methods. However, one of the most widely used methods is redirection through the server redirection response, which is the method we consider in our work. The redirection response is identified by the response status code (300-308). In this method, upon receiving the server's redirection response, the client uses the new URL provided in the Location header, and sends a new request to the new URL. The new URL that resulted from the redirection response can also perform a redirection, forming what we call a \quotes{redirection chain}. The recommended maximum number of redirections that a client can accept is five redirections~\cite{http14}. However, enforcing it may vary between client vendors.

\subsection{HTTP Headers}
HTTP headers allow clients and servers to exchange additional information in the HTTP requests and responses. HTTP headers are sent as a key/value pairs, with the header name as the key. There are many types of HTTP headers, including those for signaling and specifying security policies, which we refer to as \quotes{security headers}. In servers' security headers, the server sends one or more security headers to instruct the client to enforce a security policy. Some of the most important security headers are: 
\begin{enumerate}
	\item Content-Security-Policy (CSP)~\cite{csp19}, to restrict the sources a client can load content from, to protect against code injection attacks such as Cross-Site Scripting (XSS) attacks.
	\item Strict-Transport-Security (HSTS)~\cite{hsts19}, to inform the client to always enforce HTTPS, to provide protection against TLS layer downgrade attacks (see~\cite{alashwali18} for a taxonomy and a survey of TLS downgrade attacks). 
	\item The Secure attribute in the Set-Cookie header~\cite{cookie19}, to restrict sending the cookies over a secure HTTPS channel, to protect against leaking users' private data, and identity theft attacks.
\end{enumerate}

\subsection{Threat Model}
We assume an attacker aiming to cause an HTTPS client to connect to the requested domain in a region that provides weaker HTTPS security guarantees than those in the client's native region. The attacker's goal is to either exploit or enable a third party to exploit the \quotes{weaker} region's flaws that do not exist in the client's native region. The attacker can be represented by a man-in-the-middle, or a discriminatory server. The latter can be due to a dishonest service provider, dishonest server administrator, or a malware. More concrete attack scenarios are provided in \autoref{sec:scenario}.

\section{Empirical Study} \label{sec:empirical}
\subsection{Dataset}\label{sec:dataset}
Our dataset consists of the top \num{250000} most visited domains on the Internet, obtained from the Tranco list~\cite{pochat18}\footnote{https://tranco-list.eu/}, a configurable research oriented top domains list, hardened against manipulation attacks. We first extract the top 1 million list, then we extract the top \num{250000} from it. We choose to use the \quotes{standard} configurations of the list, which aggregates the ranks of \quotes{pay-level}\footnote{According to the list's authors, \quotes{Pay-level} domain refer to \quotes{a domain name that a consumer or business can directly register.}~\cite{pochat18}} domains, which do not contain subdomains, from four widely used top domains lists: Alexa~\cite{alexa19}, Umbrella~\cite{umbrella19}, Majestic~\cite{majestic19}, and Quantcast~\cite{quantcast19}, over the past 30 days. We retrieved the list on the 10\textsuperscript{th} of April, 2019. In our top \num{250000} domains list, the majority of domains have generic TLDs in the form of \quotes{domain.TLD}. However, our dataset contains 9.19\% domains with \quotes{multi-level} TLDs, e.g. ccTLDs, which we identify by searching for domains that have more than one dot \quotes{.}. We also check our list against Google's Safe Browsing database by setting up a local server using Google's open source Safe Browsing Go implementation~\cite{safebrowsing19}, which receives updates frequently. We query Google's Safe Browsing database using the API v4 threatMatches using an HTTP POST request to the local server. We made the query on the 27\textsuperscript{th} of June, 2019. We found a total of 250 malicious distinct domains in our dataset. The threat types, based on Google classifier, are: Malware (25 domains), unwanted\_software (29), and social_engineering (197) (note: there is one domain that appeared in two threat types). In our results in the coming sections, we exclude known malicious domains.
\subsection{Setup}\label{sec:setup}
To conduct our experiment, we create five machines (clients), configured to be identical from all aspects (hardware, software, and configurations), physically located at five different geographic locations, rented from a cloud provider (Amazon Elastic Compute Cloud (EC2)~\cite{amazonec2}). We considered choosing locations that are geographically, economically, and politically distant. Therefore, we select a country from 5 different continents. The locations we choose are (continent, country): Australia, Australia (AU); South America, Brazil (BR); Asia, India (IN); Europe, United Kingdom (UK); North America, United States (US). Note that our choices are bound by the region's availability and requirements at Amazon EC2. For example, China would be an interesting region to have, however, it has different requirements such as a valid business license from the Chinese government, which we do not have. We first launch the first client in the first region using Ubuntu 18.04 Operating System (OS) and OpenSSL 1.1.0g and an updated Certificate Authority (CA) store, and we install and configure the software we need in our experiment including the scanners and their libraries. We then create an image of the first client's disk. After that, we create the remaining four clients with the same hardware specifications, running a copy of the first client's image.
\subsection{Data Collection}\label{sec:collection}
To collect data, we run two scans: a redirection scan followed by a TLS scan. The redirection scan collects data at the application layer such as the servers' response headers and the URL redirection chains. The TLS scan collects data at the transport layer such as the servers' selected protocol versions, ciphersuites, and TLS certificates. To run the scans, we use a mixture of existing open source and those developed in-house tools.\par 

First, for the redirection scan, we develop our own redirection scanner using the Python's Requests 2.21.0 library \cite{requests19}, which allows us to automate sending HTTPS requests, collect response URLs, redirection chains (if exist), and the full content of response headers. Our scanner utilises a Transport Adapter that is configured with Chrome's latest version pre-TLS~1.3 ciphersuites. The ciphersuites list at this stage is not sensitive as we do not collect transport layer data at the redirection scan. However, the default ciphersuites list that is shipped with the Requests library is much longer than Chrome's list. Therefore, using Chrome's ciphersuites list has a performance advantage and can maximise the response rate as it contains the most widely supported ciphersuites. We disable the certificate validation as we use the redirection scanner to collect application layer data regardless of the certificate status. \par 

For the TLS scan, we use the tls-scan tool, an open source fast TLS scanner \cite{tls-scan16}. We customise the tls-scan client to utilise the OpenSSL 1.1.1g which supports versions from TLS~1.0 to TLS~1.3, and to offer Google Chrome's latest version\footnote{As of April 2019, version 74.0.3729.108.} ciphersuites, which adds support for three new TLS~1.3 ciphersuites besides Chrome's pre-TLS~1.3 ciphersuites.\par 

After setting up the experiment clients, from each region, on the same day, we first run the redirection scan which gathers application layer data. The redirection scanner takes the domain names from our dataset as input. After that, for each domain, the scanner conducts an HTTPS GET request, collects the final response URL, and the URL redirection chain (if exists) starting from the requested URL until the one before the final response URL. In addition, we collect the response status codes and the full content of the response headers. After finishing the redirection scan, in each region, for each request made, we extract the domain (including the TLD and subdomains if exist) of the HTTPS final response, to use it as an input for the TLS scan phase. Note the difference in the input domains between the redirection scan and the TLS scan. The redirection scanner makes requests for the same domains in all regions. For example, if the dataset contains \texttt{example.com}, clients in all regions make requests to \texttt{example.com}. However, since our redirection scanner follows redirections, the response domain in the redirection scan may differ based on the region, e.g. the US-based client receives \texttt{us.example.com} while the UK-based client receives \texttt{uk.example.com}. The TLS scanner takes the final response distinct domains as input, and does not perform redirection as these are the final response domains based on the redirection scan results. The TLS scan results reflect the state of the final domain. We conduct both scans between the 11\textsuperscript{th} and 20\textsuperscript{th} April, 2019. However, requests for any set of domains from the five regions are sent on the same day. Finally, after the data is collected, we load and analyse them using MySQL database and queries. We describe the data analysis methodology in more detail next.

\subsection{Ethical Considerations}\label{sec:ethical}
Our experiment is in compliance with the ethical considerations in carrying out
measurement studies. First, we do not collect any private data. The data we
collect are public meta data that are provided by websites to any client such as web browsers. Second, we do not perform so many handshakes as to exhaust any single server. Our clients' handshakes can by no means be classified as a Denial of Service (DoS) attack. 

\subsection{Data Analysis}\label{sec:analysis}
To analyse data, we define HTTPS weakness indicators based on well-known security guidelines and recommendations that are adopted by secure HTTPS clients and servers. For example, using HTTPS not HTTP, using security headers such as HSTS and CSP. In addition, we define some TLS weakness indicators based on recommendations such as NIST's recommendations on using the latest TLS version and the secure ciphersuites that provide strong guarantees such as FS and AE~\cite{nist18}. We consider the security of servers' response from three vectors: 
\begin{enumerate}
	\item URLs security. 
	\item Security headers. 
	\item TLS security. 
\end{enumerate} 

Then, to find inconsistent HTTPS cases, we count domains that satisfy a weakness indicator in \textit{\textbf{some but not all}} of their responses to our clients in the five regions. In what follows, we define the weakness indicators in each vector. 

\subsubsection{URLs Security Weakness Indicators} \label{sec:analysisurls}
	\begin{itemize}
		\item \textbf{plain-HTTP (final/intermediate) URL:} This weakness indicator is satisfied if the server's response URL is using the plain-HTTP protocol. Needless to say, plain-HTTP does not give the security guarantees of HTTPS. We analyse the server's response URL from two perspectives: the response's final (landing) URL, and the response's intermediate URL(s) that appear in the redirection chain, if any. Our clients' requests are initiated using the HTTPS protocol, but may receive plain-HTTP in intermediate or final responses, depending on the server's response.  

		\item \textbf{Incompatible (final/intermediate) domain:} This weakness indicator is satisfied if the server's response URL contains a domain that is different from the requested domain. By  \quotes{incompatible} we mean unequal, different, or inconsistent. However, we avoid reusing the term \quotes{inconsistent} to avoid ambiguity as we already use the term \quotes{inconsistent} to describe unequal response for requests from the five regions with respect to the defined weakness indicators including the \quotes{incompatible (final/intermediate) domain} indicator. Similar to the previous indicator, we analyse the server's response domain from two perspectives: the response's final (landing) URL, and the response's intermediate URL(s) that appear in the redirection chain, if any. To examine the requested domains against the final/intermediate response's domain, we extract the \quotes{pay-level} domain (a.k.a. main-domain or base-domain) of the request and response domains using the tldextract Python library~\cite{kurkowski19}, which identify the pay-level domain by maintaining an updated list of all known public TLDs obtained from Mozilla's public suffix list~\cite{mozilla19-publicsuffix}. Then, we compare them using a case-insensitive equality check. If they are not equal, they are classified as incompatible. For the intermediate domains check, since a response may contain multiple intermediate URLs, this weakness indicator is satisfied if one or more intermediate domains are incompatible with the requested domain.   
	\end{itemize}	

The compatibility check is implemented in Python. The results are then stored and queried using MySQL.  
 
\subsubsection{Security Headers Weakness Indicators} \label{sec:analysisheaders}
\begin{itemize}
	\item \textbf{No-HSTS:} This weakness indicator is satisfied if the Strict-Transport-Security header is absent from the server's response.  
	\item \textbf{No-CSP:} Similar to the previous indicator but for the Content-Security-Policy header.
	\item \textbf{No-\quotes{Secure}-Set-Cookie:} This indicator is satisfied if the Set-Cookie header is sent in the servers' responses to the five regions' client requests to a particular domain, and one or more cookie values in the Set-Cookie header do not contain the Secure attribute value.
\end{itemize}

The headers' presence check is implemented in Python by parsing the headers which are loaded as a dictionary of key/value objects,  where the header name is the key. The results are then stored and queried in MySQL.  
\subsubsection{TLS Weakness Indicators} \label{sec:analysistls}
\begin{itemize}
	\item \textbf{Version \textless TLS~1.3:} This indicator is satisfied if the server selects a TLS version less than TLS~1.3. TLS~1.3 is the latest version of the TLS protocol. Since it was standardised in August 2018 and our scan is in April 2019, the resulting inconsistencies against TLS~1.3 can be influenced by its recency. For this reason, we also examine \quotes{Version \textless TLS~1.2} weakness indicator, to have a balanced perspective.   

	\item \textbf{Version \textless TLS~1.2:} This indicator is satisfied if the server selects a TLS version less than TLS~1.2. Versions less than TLS~1.2 are officially weak and should not be used today. This indicator is negated if the server selects version TLS~1.2 or TLS~1.3.

	\item \textbf{Non-FS:} This indicator is satisfied if the server selects a non-FS-ciphersuite. We define\footnote{Our ciphersuites' definitions are based on Chrome's TLS configurations.} non-FS-ciphersuites as those that do not support the Elliptic-Curve Diffie-Hellman (ECDHE) key-exchange, and also are not negotiated with version TLS~1.3 since TLS~1.3 enforces FS by design in separate extensions. 

	\item \textbf{Non-FS+Non-AE:} This indicator is satisfied if the server selects a non-FS and a non-AE ciphersuite (non-FS+non-AE-ciphersuite), i.e. neither provide FS nor AE. We define non-FS+non-AE ciphersuites as those that do not support ECDHE, are not negotiated with version TLS~1.3, do not support the ChaCha20 symmetric encryption, and do not support the GCM symmetric encryption mode. This indicator is negated if the server selects either a FS-ciphersuite, or AE-ciphersuite, or FS+AE-ciphersuite. 
	
	\item \textbf{Expired Certificate:} This indicator is satisfied if the server provides an expired certificate. The certificate expiration date is checked against the scan date using the tls-scan client \cite{tls-scan16}. 

	\item \textbf{Invalid Host Name:} This indicator is satisfied if the server provides an invalid host name in the certificate. The validation is based on the tls-scan client using the X509\_check\_host function in the OpenSSL library~\cite{openssl19}. It checks if the certificate's Subject Common Name (CN) or Subject Alternative Name (SAN) matches the specified host name.
\end{itemize}

\section{Results} \label{sec:results}
In this section, we first provide a general overview of the response data. Then, we provide the results of our inconsistency analysis against each weakness indicator, which we defined in \autoref{sec:analysis}. In this section, we are concerned about inconsistencies. That is, we quantify domains that satisfy a weakness indicator in \textit{\textbf{some but not all}} of their responses to our clients' requests from the five regions. We do not report domains that consistently satisfy a weakness indicator in their responses to the clients in the five regions. 

\subsection{Responses}\label{sec:responses}
\begin{table*}[!tp]
	\centering
	\caption{Summary of the redirection scan responses. The percentages of the responses \quotes{(Any/Valid) HTTP(S) Final Responses} are computed over the \num{250000} input domains. Each indentation in a row means that the percentages of that row are computed from the previous indentation level.}
	\label{tab:responses} 
		\begin{tabular}{lcr@{\hspace{5pt}}r}
			\toprule  
			Response Type& Status Code  &
			\multicolumn{2}{c}{Joint Responses}\\  
					
			\midrule 
			\quad Any HTTP(S) Final Response & Any &
			\num{187902}&(\fpeval{round((187902/250000)*100,2)}\%) \\
			
			\quad \quad Any consistent plain-HTTP Final Response & Any & 
			\num{10320}&(\fpeval{round((10320/187902)*100,2)}\%) \\
			
			\quad \quad Any consistent HTTPS Final Response & Any & 
			\num{176936}&(\fpeval{round((176936/187902)*100,2)}\%) \\
			
			\quad \quad \quad Any consistent HTTPS with Redirections $>$ 0  & Any & \num{86123}&(\fpeval{round((86123/176936)*100,2)}\%) \\
			
			\quad \quad \quad \quad Any consistent HTTPS with plain-HTTP Inter. Response & Any &
			\num{12198}&(\fpeval{round((12198/86123)*100,2)}\%) \\
			
			
			\midrule
			\rowcolor{lightYellow}
			
			\quad Valid HTTP(S) Final Response& 200 &
			\num{163235}&(\fpeval{round((163235/250000)*100,2)}\%)\\
			
			\quad \quad Valid consistent plain-HTTP Final Response & 200 & \num{9242}&(\fpeval{round((9242/163235)*100,2)}\%)\\
			
			\rowcolor{lightYellow}
			\quad \quad Valid consistent HTTPS Final Response &200& \num{153761}&(\fpeval{round((153761/163235)*100,2)}\%) \\
			
			\quad \quad \quad Valid consistent HTTPS with Redirections $>$ 0  & 200& \num{81019}&(\fpeval{round((81019/153761)*100,2)}\%) \\
			
			\quad \quad \quad \quad Valid consistent HTTPS with plain-HTTP Inter. Response & 200& 
			\num{11873}&(\fpeval{round((11873/81019)*100,2)}\%) \\
			\bottomrule
		\end{tabular}
\end{table*}


\autoref{tab:responses} provides a summary of the \quotes{joint} response data. That is, a joint response is counted if we receive responses for requests from the 5 regions. We follow a best effort approach in data collections. That is, we do not investigate the reasons of failed responses in one or more regions, which can be for various reasons, and we count as a joint fail. However, the failure rate affects the collected data size, but not the validity of the collected data. We compile the data twice: first, without considering the response status code, i.e. any code (see the first half of \autoref{tab:responses}), which we prefix them by \quotes{Any}. Second, when considering only those responses that provide the \quotes{200 OK} status code, which indicates that the HTTP request has succeeded (see the second half of \autoref{tab:responses}), which we prefix them by \quotes{Valid}. We then dissect the overall HTTP(S) responses in both categories, i.e. the (Any/Valid) categories, into two sub-categories: \quotes{(Any/Valid) plain-HTTP Final response}, which denotes responses with the insecure plain-text HTTP protocol (without TLS), and \quotes{(Any/Valid) HTTPS Final Response}, which denotes responses with the secure HTTPS protocol. From the \quotes{(Any/Valid) HTTPS Final Responses}, we  count those with \quotes{(Any/Valid) Redirections \textgreater 0}. From those, we count the redirections that contain one or more \quotes{(Any/Valid) plain-HTTP Intermediate Response}. As depicted in \autoref{tab:responses}, the \quotes{Joint Response} column shows the number of cases that received responses from the five regions. Finally, the percentages of the overall responses \quotes{(Any/Valid) HTTP(S) Final Responses} are computed over the dataset size (\num{250000}). However, each indented row in \autoref{tab:responses} means that the percentages of that row are computed over the previous indentation level. \par 

In the remaining sections, we base our analysis on the valid  HTTP(S) and the valid HTTPS final responses (see the highlighted rows in the second half of \autoref{tab:responses}).

\subsection{HTTPS Security Inconsistencies}\label{sec:inconsistencies}
We now summarise the HTTPS security inconsistent cases that we find from three vectors:
\begin{enumerate}
	\item URLs security.
	\item Security headers.
	\item TLS security.
\end{enumerate}

We now divide this section into three subsections according to the three vectors listed above as follows:
  
\subsubsection{URLs Security}\label{sec:urls}
In this section, we consider the valid HTTP(S) final responses (the highlighted row labeled \quotes{Valid HTTP(S) Final Response} in \autoref{tab:responses}) as we require both plain-HTTP as well as HTTPS responses to be analysed. The results of our inconsistency analysis of URLs security are summarised in \autoref{tab:url}. The following subsections are divided according to the security properties related to the \quotes{URLs security} vector (see~\autoref{tab:vectors}), using the weakness indicators that we defined in our data analysis in~\autoref{sec:analysisurls} under the \quotes{URLs security} vector. \par 

\begin{table}[!tp]
	\centering
	\caption{Summary of the URLs security inconsistencies. The \quotes{Inconsistent} column shows the overall inconsistent cases against the weakness indicator. The percentages are computed over the \num{163235} valid joint HTTP(S) responses.}
	\label{tab:url} 	
		\begin{tabular}{lr@{\hspace{5pt}}r}
			\toprule 
			Weakness Indicator & \multicolumn{2}{c}{Inconsistent} \\
			\midrule 
			plain-HTTP Final URL & 
			\num{232} & (\fpeval{round((232/163235)*100,2)}\%)\\
			
			plain-HTTP Inter. URL & 
			\num{501} & (\fpeval{round((501/163235)*100,2)}\%) \\
			
			\midrule 
			Incompatible Final URL & 
			\num{280} & (\fpeval{round((280/163235)*100,2)}\%) \\
			
			Incompatible Inter. URL & 
			\num{205} & (\fpeval{round((205/163235)*100,2)}\%)\\
			\bottomrule
		\end{tabular}
\end{table}

\begin{figure}[tp!]
	\centering
	\begin{subfigure}[b]{0.9\columnwidth}
		\includegraphics[width=\linewidth]{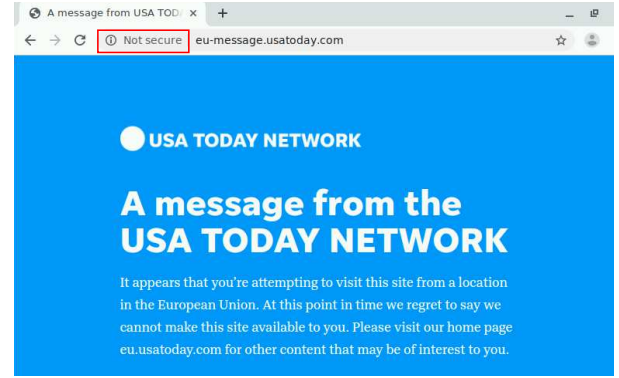}
		\caption{\texttt{gannett.com} over plain-HTTP in the UK due to blocking.}
		\vspace{5pt}
	\end{subfigure}
	
	\begin{subfigure}[b]{0.9\columnwidth}	
		\vspace{5pt}
		\includegraphics[width=\linewidth]{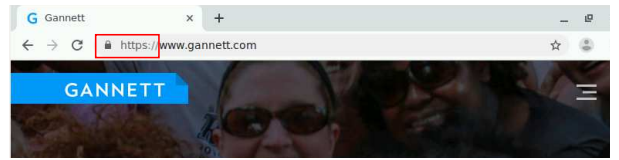}
		\caption{\texttt{gannett.com} over HTTPS in the US.}
	\end{subfigure}
	\vspace{10pt}
	\caption{The case of \texttt{gannett.com} in the UK vs. the US.}
	\label{fig:gannett}
\end{figure}

\paragraph{HTTPS in Final URLs} 
Using the \textbf{plain-HTTP final URL} weakness indicator, we find 232 inconsistent cases for domains that send their final responses in plain-HTTP URL to some but not all of the five regions. Needless to say, domains that send plain-HTTP responses are prone to impersonation attacks due to lack of authentication, data manipulation during transmission due to lack of integrity, and espionage, e.g. to steal users' credentials, due to a lack of secrecy (encryption). To understand the reasons for inconsistent plain-HTTP responses, we manually inspect the plain-HTTP final URLs in each region, within around two weeks after completing the scans. Through the Remote Desktop Protocol (RDP), we connect to the remote client, and from there we manually visit each website that resulted in plain-HTTP response, using the Firefox web browser. We identify several reasons for inconsistent plain-HTTP responses. We find that blocking pages play a considerable role in such inconsistencies, where these domains provide a valid plain-HTTP response (200 OK) status code, but display a plain-HTTP blocking page to inform the user that the requested domain (website) is not accessible in the client's region. See \autoref{fig:sendspace} for \texttt{sendspace.com} case, and \autoref{fig:gannett} for \texttt{gannett.com} case. The latter is an American media company which owns the \quotes{US Today} newspaper. Out of the active websites at the time of the manual inspection, we identify blocking pages in (note: the total numbers provided here are excluding domains that did not respond to the manual inspection): 19/80 (23.75\%) in AU, 20/98 (20.41\%) in BR, 42/132 (31.82\%) in IN, 28/93 (30.11\%) in the UK, and 4/71 (5.63\%) in the US. Clearly out of the five regions, IN has the highest percentage of blocking pages, followed by the UK, while the US has the lowest percentage. There are various reasons for regional blocking as also noted in~\cite{afroz18}, including, but not limited to: government orders (observed mostly in IN. See \autoref{fig:sendspace} for example), GDPR compliance (mostly in the UK), business unavailability (e.g. a company does not offer its products to some regions). One might argue that blocking pages do not contain sensitive data and therefore there is little or no impact from the absence of HTTPS on these pages. This argument is true with respect to secrecy. However, using TLS in HTTP (i.e. HTTPS) is meant to provide secrecy, integrity, and authentication to prevent \quotes{eavesdropping, tampering, and message forgery}~\cite{rescorla18tls13}. Therefore, the absence of HTTPS even in blocking pages is harmful. It makes the page's authenticity and integrity questionable. That is, users can not trust whether the plain-HTTP response is a genuine blocking page or a faked response, e.g. to deny users from reaching the genuine website to cause losses to it.

\begin{figure}[tp!]
	\centering
	\begin{subfigure}[b]{0.9\columnwidth}
		\includegraphics[width=\linewidth]{fig/govblocking/\detokenize{in_sendspace}}
		\caption{\texttt{sendspace.com} over plain-HTTP in the IN due to blocking.}
		\vspace{5pt}
	\end{subfigure}
	
	\begin{subfigure}[b]{0.9\columnwidth}	
		\vspace{5pt}
		\includegraphics[width=\linewidth]{fig/govblocking/\detokenize{us_sendspace}}
		\caption{\texttt{sendspace.com} over HTTPS in the US.}
	\end{subfigure}
	\vspace{10pt}
	\caption{The case of \texttt{sendspace.com} in IN vs. the US.}
	\label{fig:sendspace}
\end{figure} 


\begin{figure}[tp!]
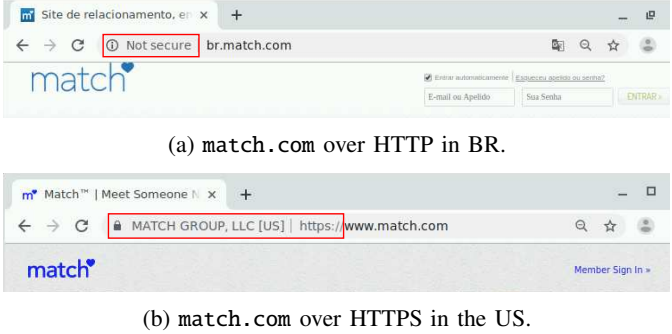

	\centering
	\begin{subfigure}[b]{\columnwidth}
		\includegraphics[width=\textwidth]{fig/http/match/\detokenize{br_match}}
		\caption{\texttt{match.com} over HTTP in BR.}
		\label{fig:brmatch}
	\end{subfigure}
	
	\begin{subfigure}[b]{\columnwidth}	
		\vspace{5pt}
		\includegraphics[width=1\linewidth]{fig/http/match/\detokenize{us_match}}
		\caption{\texttt{match.com} over HTTPS in the US.}
		\label{fig:usmatch}
	\end{subfigure}
	\vspace{5pt}
	\caption{The case of \texttt{match.com} in BR vs. the US.}
	\label{fig:match}
\end{figure}

Apart from regional blocking, there are indeed inconsistent cases where the same domain is deployed in plain-HTTP in some but not all regions. For example, the case of \texttt{match.com}, a high profile dating website, if visited from BR, the client is redirected to the plain-HTTP domain \texttt{br.match.com}, while all the other examined regions, such as the US, are provided with HTTPS. See \autoref{fig:match} for illustration. Interestingly, we revisited \texttt{match.com} from BR around four months after our initial observation. We find that it stopped redirecting to \texttt{br.match.com}, and when we manually visit \texttt{br.match.com} directly, we notice that it is using HTTPS, with a certificate starting from March 8, 2019, around one month before our scan. This indicates that the insecurity of \texttt{br.match.com} that we initially observed is due to insecure redirection, which redirects to the insecure plain-HTTP version of the website. From this case we find that visiting the regional domains manually, e.g. \texttt{br.match.com} can be more secure than visiting the generic domain, e.g. \texttt{match.com} that redirects to the regional domain, to avoid potential redirection misconfigurations.     

\begin{figure}[tp!]
	\centering
	\begin{subfigure}[b]{\columnwidth}
		\includegraphics[width=\columnwidth]{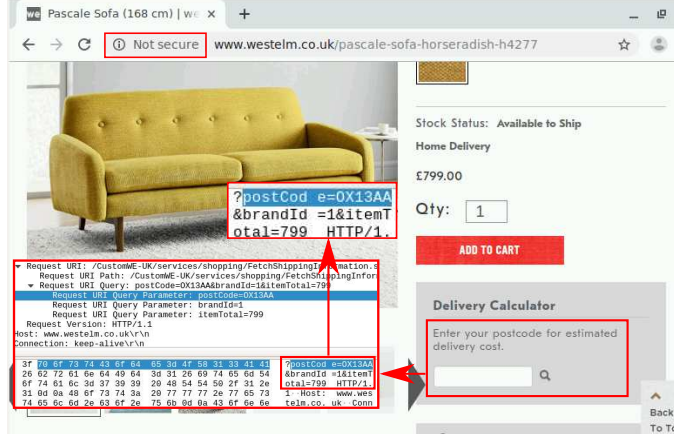}
		\caption{\texttt{westelm.com} browsing over HTTP in the UK. The image in the border shows some personal data such as postcode and product prices are sent in clear-text.}
		\label{fig:ukwestelm}
	\end{subfigure}
	
	\begin{subfigure}[b]{\columnwidth}	
		\vspace{5pt}
		\includegraphics[width=1\linewidth]{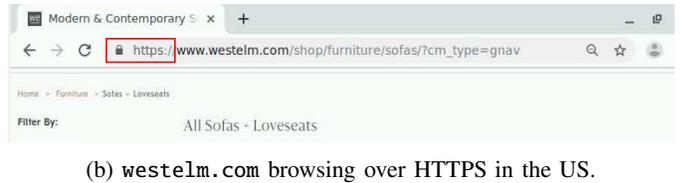}
		\caption{\texttt{westelm.com} browsing over HTTPS in the US.}
		\label{fig:uswestelm}
	\end{subfigure}
	\vspace{5pt}
	\caption{The case of \texttt{westelm.com} in the UK vs. the US.}
	\label{fig:westelm}
\end{figure}

We also identify inconsistent cases of plain-HTTP due to partial HTTPS in some but not all regions. By partial HTTPS we refer to websites deploying HTTPS partially, e.g. in some pages such as login pages, while leaving the rest of the pages sent over plain-HTTP, which endangers users' privacy. This is the case we find in \texttt{westelm.com} which deploys partial HTTPS in the UK, while deploying full HTTPS in other regions such as the US. See \autoref{fig:westelm}, which shows that some data such as users' postcode and item prices are sent in plaintext for the UK clients. We have emailed the customer service about this weakness, and we offered clarification and help, but we have not heard back yet.\par 


Despite the fact that the user credentials of the weak and strong domains in the aforementioned two examples: the US's \texttt{match.com} versus the BR's \texttt{br.match.com}, and the US's \texttt{westelm.com} versus the UK's \texttt{westelm.co.uk}, are not shared, and each region requires a separate account (we tested this manually), plain-HTTP final URL inconsistencies are still dangerous for several reasons. First, it provides degraded security for some regions' users. Second, high profile domains provide a false sense of security. For example, a user in one region that used to receive strong HTTPS security guarantees, will expect the same security guarantees if the user moved or traveled to another region. Third, the scenario of shared user credentials between different regions is not unusual, and we have found it used in \texttt{amazon.com} for example, where we can use a single credential among various regional domains such as \texttt{amazon.com} for the US users and \texttt{amazon.co.uk} for the UK users. This can enable an attacker who succeeds in redirecting a user from a secure version of the website in one region to an insecure one in another region, from compromising the user's credentials while using the insecure domain. \par 

\paragraph{HTTPS in Intermediate URLs}
Using the \textbf{plain-HTTP intermediate URLs} weakness indicator, we find 501 inconsistent cases, where one or more plain-HTTP intermediate URLs are found in some but not all of the responses to the five regions. Note that the results in this indicator are computed out of HTTP(S) responses. However, if we compile the plain-HTTP intermediate URLs inconsistent cases out of the HTTPS final responses only, the number of inconsistent plain-HTTP intermediate URLs drops to 344, which means that 31.34\% of plain-HTTP inconsistent intermediate URLs cases contains plain-HTTP final URLs in one or more regions. Moreover, there are \num{105} (20.96\%) domains of the plain-HTTP intermediate URLs inconsistent cases that intersect with the domains of plain-HTTP final URLs inconsistent cases. Unlike final URLs which are normally visible to the user, e.g. through the URL bar in web browsers, intermediate URLs in redirection chains are invisible to the user, which make inconsistencies in plain-HTTP intermediate URLs worse than those in final URLs. plain-HTTP intermediate URLs can enable a man-in-the-middle attacker to impersonate an intermediate domain, and redirect the user to a malicious, e.g. a phishing website. \par  

\begin{table*}[!tp]
	\centering
	\caption{Examples of incompatible final domains and the regions they are found in.}
	\label{tab:incompatibleexamples} 
	\begin{adjustbox}{width=0.9\textwidth}	
		\begin{tabular}{lll}
			\toprule 		
			Region & Initial Domain & Final Domain\\
			\midrule 	
			AU	& \texttt{https://bbcchannels.com} & \texttt{https://www.bbcaustralia.com}\\
			\hline
			UK & \texttt{https://aa.com} & \texttt{https://www.americanairlines.co.uk/homePage.do?locale=en_GB}\\
			\hline 
			UK & \texttt{https://live2all.com} & \texttt{https://www.livetotal.net} \\
			\hline 
			UK & \texttt{https://cartoonnetworkasia.com} & \texttt{https://www.cartoonnetworkeurope.com} \\
			\hline 
			US	& \texttt{https://hoteis.com} & \texttt{https://www.hotels.com/?pos=HCOM\_US\&locale=en\_US} \\
			\bottomrule  
		\end{tabular}
	\end{adjustbox}
\end{table*}


\paragraph{Compatible Domains in Final URLs} 
Using the \textbf{Incompatible Final URLs} weakness indicator, we find \num{280} inconsistent cases for domains that send incompatible final URLs in their responses to some but not all regions. We consider incompatible final URLs as a weakness indicator as they contradict the common security advice to users which recommends visually checking the compatibility of the requested domain (e.g. in e-mail links) against the displayed response domain, and suspect response domains that are incompatible with the requested domain. Although HTTP(S) response domains can be incompatible with the requested domain for benign reasons (e.g. a different brand name for a company), having incompatible response domains in some but not all of the regions is particularly suspicious. \autoref{tab:incompatibleexamples} provides few examples which we also cross-checked with the Chrome browser and found the same behavior that our dataset indicates. The examples also illustrate the difficulty of judging whether an incompatible final domain name is benign or not. For example, in the case of \texttt{hoteis.com}, the response domain \texttt{hotels.com} differs from the requested domain only in one letter, which is a common phishing technique. It also shows the lack of standard domain name format for regional domain names. While many domains tend to change either the subdomain or the TLD for regional domains, we found several cases of completely different domain names with regional indication. For example, requesting \texttt{bbcchannels.com} from the AU is redirected to  \texttt{www.bbcaustralia.com}, a whole different domain name. \autoref{tab:incompatibletefal} shows an example for \texttt{tefal.com} which illustrates the incompatible final URLs inconsistency. That is the final domain differs from the requested domain in some but not all of the regions: the US and BR final domains are different than the requested domain, while the AU, IN and UK final domains are identical to the requested one (except the TLDs which are tolerated). The lack of standard regional domain names hardens the verifiability of the requested domain, and opens a door for phishing attacks when users benignly believe that the new domain name is a result of regional redirection. 
 		
\begin{table}[!tp]
	\centering
	\caption{The \texttt{tefal.com} case of incompatible final response URLs found in BR and US regions.}
	\label{tab:incompatibletefal} 
	\begin{tabular}{lll}
		\toprule 		
		Region & Requested Domain & Final Domain\\
		\midrule 
		AU & \texttt{https://tefal.com} & \texttt{https://www.tefal.com.au} \\
		\rowcolor{lightYellow}
		BR	& \texttt{https://tefal.com} & \texttt{https://www.arno.com.br}\\
		IN & \texttt{https://tefal.com} & \texttt{https://www.tefal.in} \\
		UK& \texttt{https://tefal.com} & \texttt{https://www.tefal.co.uk} \\
		\rowcolor{lightYellow}
		US & \texttt{https://tefal.com} & \texttt{https://www.t-falusa.com} \\
		\bottomrule
	\end{tabular}		
\end{table}

\paragraph{Compatible Domains in Intermediate URLs}
Using the \textbf{Incompatible Intermediate URLs} weakness indicator, we find \num{205} inconsistent cases for domains that send one or more incompatible intermediate URLs in their responses to some but not all regions. Note that we count the incompatible intermediate URLs inconsistencies regardless of the status of the final URLs. However, there are \num{100} (48.78\%) of the domains in the incompatible intermediate URLs inconsistent cases that intersect with the domains of incompatible final URLs inconsistent cases. Unlike final URLs, intermediate URLs are requested silently in the background and are invisible to the user, which makes incompatible intermediate URLs even more suspicious than incompatible final URLs.

\subsubsection{Security Headers}\label{sec:headers}
The results of our analysis of security headers' inconsistencies are computed over the valid HTTPS responses only as measuring security headers' inconsistencies is more meaningful in HTTPS connections. For the cookies, we check the consistency of the Secure attribute among HTTPS responses that have the Set-Cookie header in all the five responses to our clients' requests in the five regions. \autoref{tab:headers} summarises the results. The following subsections are divided according to the security properties related to the \quotes{security headers} vector (see~\autoref{tab:vectors}), using the weakness indicators that we defined in our data analysis in~\autoref{sec:analysisurls} under the \quotes{security headers} vector.

\begin{table}[!tp]
	\centering
	\caption{Summary of the security headers' inconsistencies.  The \quotes{Inconsistent} column shows the overall inconsistent cases against the weakness indicator. The No-HSTS and No-CSP percentages are computed over the \num{153761} valid joint HTTPS responses. The No-Secure in Set-Cookie percentage is computed over the \num{77568} valid joint HTTPS responses that contain Set-Cookie headers.}
	\label{tab:headers} 	
	\begin{tabular}{lr@{\hspace{5pt}}r}
		\toprule  
		Weakness Indicator & \multicolumn{2}{c}{Inconsistent}\\ 
		\midrule 
		No-HSTS  
		& \num{517} &(\fpeval{round((517/153761)*100,2)}\%)\\
		
		No-CSP  
		& \num{481} &(\fpeval{round((481/153761)*100,2)}\%) \\
				
		No-Secure in Set-Cookie  
		& \num{223} &(\fpeval{round((223/77568)*100,2)}\%) \\
		\bottomrule
	\end{tabular}
\end{table}

\paragraph{HSTS headers}
Using the \quotes{\textbf{No-HSTS}} weakness indicator, we find 517 inconsistent cases, where domains do not send the HSTS header in their responses to some but not all regions. The HSTS header has highest number of inconsistency cases we find in the security headers vector.

\paragraph{CSP headers}
Using the \quotes{\textbf{No-CSP}} weakness indicator, we find 481 inconsistent cases, where domains do not send the CSP header in their responses to some but not all regions. The CSP header has the second highest number of inconsistent cases we find in the security headers vector. 


\begin{figure}[tp!]
	\centering
	\begin{subfigure}[b]{0.9\columnwidth}
		\includegraphics[width=\linewidth]{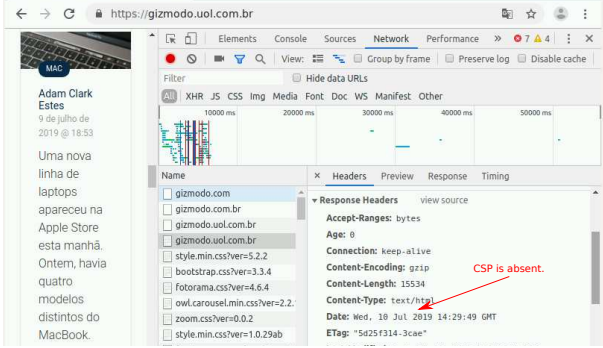}
		\caption{The CSP header is absent in \texttt{gizmodo.com} in BR.}
		\vspace{5pt}
	\end{subfigure}
	
	\begin{subfigure}[b]{0.9\columnwidth}	
		\vspace{5pt}
		\includegraphics[width=\linewidth]{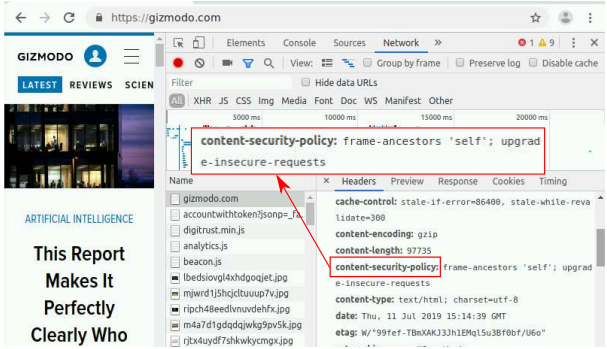}
		\caption{The CSP header is absent in \texttt{gizmodo.com} in US.}
	\end{subfigure}
	\vspace{10pt}
	\caption{The case of \texttt{gizmodo.com} in BR vs. the US.}
	\label{fig:gizmodo}
\end{figure}  

\paragraph{The Secure attribute in Set-Cookie headers}
Using the \quotes{\textbf{No-Secure in Set-Cookie}} weakness indicator, we find \num{223} inconsistent cases, where domains do not send the Secure attribute in the Set-Cookie header in their responses to some but not all regions. The Secure attribute in the Set-Cookie headers has the lowest number of inconsistent cases we find in the security headers vector.

\subsubsection{TLS Security}\label{sec:tls}
The results of our analysis of transport layer security (TLS) inconsistencies are computed over the valid HTTPS responses that also responded to the TLS scan. The results of our analysis of TLS inconsistencies are summarised in \autoref{tab:tls}. The following subsections are divided according to the security properties related to the \quotes{TLS security} vector (see~\autoref{tab:vectors}), using the weakness indicators that we defined in our data analysis in~\autoref{sec:analysisurls} under the \quotes{TLS security} vector.

\begin{table}[!tp]
	\centering
	\caption{Summary of the TLS security inconsistencies.  The \quotes{Inconsistent} column shows the overall inconsistency cases against the weakness indicator. The percentages are computed over the \num{152836} valid joint HTTPS responses that also have responses in the TLS scan.}
	\label{tab:tls} 
	\begin{tabular}{lr@{\hspace{5pt}}r}
		
		\toprule 
		
		Weakness Indicator &\multicolumn{2}{c}{Inconsistent}\\
		\midrule 
		
		Version \textless TLS~1.3  
		& \num{579} &(\fpeval{round((579/152836)*100,2)}\%)\\ 
	
		Version \textless TLS~1.2  
		& \num{54} &(\fpeval{round((54/152836)*100,2)}\%)\\ 
		
		Non-FS  
		& \num{100}&(\fpeval{round((100/152836)*100,2)}\%) \\
		
		Non-FS+Non-AE  
		& \num{71} &(\fpeval{round((71/152836)*100,2)}\%) \\
		
		Expired Cert.  
		& \num{21} &(\fpeval{round((21/152836)*100,2)}\%) \\
		
		Invalid Cert. Host Name  
		& \num{54} &(\fpeval{round((54/152836)*100,2)}\%)\\
		
		\bottomrule
	\end{tabular}
\end{table}

\paragraph{Protocol version TLS~1.3}
Using the \quotes{\textbf{Version \textless TLS~1.3}} weakness indicator, we find \num{579} inconsistent cases, where domains use version \textless TLS~1.3 in their responses to some but not all regions. This is the highest number of inconsistent cases we find in the TLS security vector. Versions \textless TLS~1.3 are prone to various attacks including several cases of downgrade attacks (see~\cite{alashwali18} for a background and a survey of downgrade attacks).  

\paragraph{Protocol version TLS~1.2 or higher}
Using the \quotes{\textbf{Version \textless TLS~1.2}} weakness indicator, we find \num{54} inconsistent cases, where domains use version \textless TLS~1.2 in their responses to some but not all regions. TLS versions \textless TLS~1.2 do not support AE, hence, are prone to attacks over the MAC-then-encrypt schemes.
	
\paragraph{The Forward Secrecy (FS) property}
Using the \quotes{\textbf{Non-FS}} weakness indicator, we find \num{100} inconsistent cases, where domains use Non-FS ciphersuites in their responses to some but not all regions. Websites that select Non-FS ciphersuites endanger users' data for future decryption as non-FS ciphersuites do not provide protection to past session keys if the servers' long-term key is compromised at some point in time.

\paragraph{The FS and Authenticate Encryption (AE) properties}
Using the \quotes{\textbf{Non-FS+Non-AE}} weakness indicator, we find \num{71} inconsistent cases, where domains use Non-FS+Non-AE ciphersuites in their responses to some but not all regions. Non-AE ciphersuites are prone to some attacks over the MAC-then-encrypt schemes as shown in~\cite{vaudenay02}\cite{alfardan13}.

\paragraph{Non-expired certificates}
Using the \quotes{\textbf{Expired Cert.}} weakness indicator, we find only \num{21} inconsistent cases\footnote{We exclude six inconsistent cases from the result of our query for inconsistent cases using the \quotes{\textbf{Expired Cert.}} weakness indicator. In these cases, the inconsistencies arise from the fact that these certificates expired during the scan, where some clients connected to the server before the certificate's expiration while others connected to it after the certificate's expiration. Since the reason is related to the scan time, we decided to exclude them.}, where domains send expired certificates in their responses to some but not all regions. This is the lowest inconsistency in the TLS security vector. 

\paragraph{Valid hostnames in certificates}
Using the \quotes{\textbf{Invalid Host Name}} weakness indicator, we find \num{54} inconsistent cases, where domains send invalid hostnames in certificates in their responses to some but not all of regions. 

\subsubsection{Relationship to URLs and IPs Diversity, and to Redirection Presence}\label{sec:relationship}
We also check whether HTTPS security inconsistencies have a relationship to URLs and IPs diversity, and to the presence of redirections. To this end, we define three new criteria: \begin{enumerate}
	\item Diverse URLs: is satisfied if the final URLs of a server's responses to the five regions clients' requests to a particular domain are not equal. We compare the final responses' URLs as is, including the protocol scheme part (i.e. http(s)://).  
	\item Diverse IPs: is satisfied if the IPs of a server's responses to the clients' requests to a particular domain are not equal. 
	\item Redirections \textgreater 0: is satisfied if a server's responses to the clients' requests to a particular domain contain one or more redirections.
\end{enumerate}

After that, for each criterion, we compute the number of domains that satisfy the criterion under two conditions: \begin{enumerate} 
	
	\item \textbf{Inconsistent} HTTPS responses that have some but not all responses satisfy a weakness indicator denoted by \quotes{Inconsistent} in \autoref{tab:relationship}. For example, the \quotes{expired cert.} weakness indicator is satisfied if some but not all responses have expired certificate, and \quotes{version \textless TLS~1.2} if some but not all responses have version \textless TLS~1.2.
	
	\item \textbf{Secure} HTTPS responses that \textit{do not} satisfy the weakness indicator (i.e. satisfy the negation of the weakness indicator that is specified in the second column in \autoref{tab:relationship}) denoted by \quotes{Secure} in \autoref{tab:relationship}. For example, in the \quotes{expired cert.} weakness indicator, the secure cases are those that receive responses with valid certificates for the five regions clients' requests, and for \quotes{Version \textless TLS~1.2} indicator, the \textbf{Secure} cases are those that receive responses with versions $\geq$ TLS~1.2 to the five regions clients' requests.\end{enumerate} 

The higher the number we obtain from this calculation, the stronger relationship to the criterion (divers URLs, diverse IPs, or redirection \textgreater 0). To compute the percentages, as depicted in \autoref{tab:relationship}, for each of the three criteria we examine, we divide the number of \quotes{Inconsistent} cases that also satisfy the examined criterion by the total number of inconsistent cases. The same is applied to the \quotes{Secure} cases. We divide the number of \quotes{Secure} cases that also satisfy the examined criterion by the total number of secure cases. It should be noted that the total number of inconsistent cases, i.e. the divisor of the \quotes{Inconsistent} columns in \autoref{tab:relationship} should be equal to the results we obtained in our inconsistency analysis in \autoref{tab:url}, \autoref{tab:headers}, and \autoref{tab:tls}, expect in the \quotes{Divers IPs} criterion where this does not hold because we compute the total inconsistent cases that responded to the TLS scan, as we only collect IPs in the TLS scan, while the inconsistent cases in \autoref{tab:url}, \autoref{tab:headers} are computed from the redirection scan.\par 

\begin{figure}[tp!]
	\centering
	\includegraphics[width=\columnwidth]{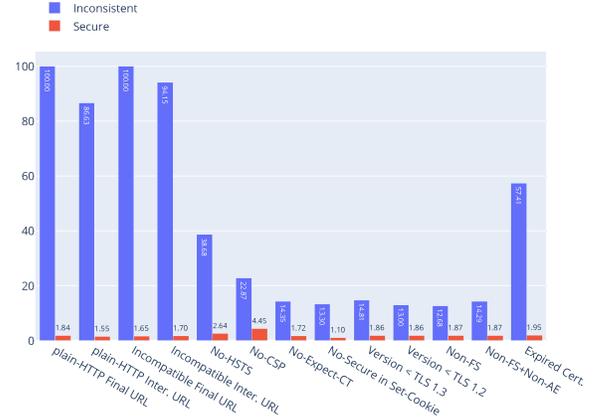}
\caption{A chart illustrating the percentages of diverse URLs in inconsistent versus secure HTTPS domains.}
\label{fig:chart}
\end{figure}

The results show that domains with inconsistent HTTPS security tend to have higher percentages of diverse URLs and IPs, and to a lesser extent redirections, than secure domains. There are few exceptions (highlighted in gray color) in the TLS weakness indicators, where inconsistent cases with respect to \quotes{Non-FS+Non-AE} indicator have less diverse IPs than those in secure cases. Similarly, inconsistent cases with respect to \quotes{Version \textless TLS~1.2} and \quotes{Expired Cert.} has less redirection \textgreater 0 than secure cases. \autoref{tab:relationship} summarises these results, and \autoref{fig:chart} demonstrates the percentages of URL diversity in inconsistent HTTPS responses compared to the secure ones. Note that inconsistent HTTPS against the \quotes{plain-HTTP Final URL} and \quotes{Incompatible Final URL} weakness indicators imply diverse URLs. This is because it is always the case that some URLs differ in the URL scheme (\quotes{https} vs. \quotes{http}) in inconsistent HTTPS cases against the \quotes{plain-HTTP Final URL} weakness indicator, which implies diverse URLs. Similarly, it is always the case that some URLs differ in the domain names in inconsistent HTTPS cases against the \quotes{Incompatible Final URL} weakness indicator. This explains the 100\% diverse URLs in the inconsistent cases against these two weakness indicators in \autoref{tab:relationship}. The same applies to the HTTPS inconsistent cases against the following weakness indicators: \quotes{plain-HTTP Final URL}, \quotes{plain-HTTP Inter. URL}, \quotes{Incompatible Final URL}, and \quotes{Incompatible Inter. URL}: they imply redirection, therefore the percentage of diverse URLs with them is 100\% as illustrated in \autoref{tab:relationship}.
 
\begin{table*}[!tp]
\centering
\caption{The percentage of diverse final URLs, diverse IPs, and redirections \textgreater 0 in two cases: inconsistent HTTPS responses that have some but not all responses satisfy a weakness indicator (Inconsistent) and consistent HTTPS responses that do not satisfy the weakness indicator (Secure). The \quotes{Diverse IPs} column results are computed over domains that responded to the TLS scan as we retrieve the IPs in the TLS scan only.}
\label{tab:relationship} 
\begin{adjustbox}{max width=\textwidth}		
	\begin{tabular}{c|lr@{\hspace{5pt}}rr@{\hspace{5pt}}rcr@{\hspace{5pt}}rr@{\hspace{5pt}}rcr@{\hspace{5pt}}rr@{\hspace{5pt}}r}
	\toprule 
	
	\multirow{2}{*}{Vector} & \multirow{2}{*}{Weakness Indicator} & \multicolumn{4}{c}{Diverse URLs} & &
	\multicolumn{4}{c}{Diverse IPs} & &
	\multicolumn{4}{c}{Redirection \textgreater 0}  \\
	\cline{3-6}\cline{8-11}\cline{13-16}
	
	& & \multicolumn{2}{c}{Inconsistent} & \multicolumn{2}{c}{Secure} & & \multicolumn{2}{c}{Inconsistent} & \multicolumn{2}{c}{Secure} & & \multicolumn{2}{c}{Inconsistent} & \multicolumn{2}{c}{Secure}\\
	\midrule 
						
	\multirow{4}{*}{\rotatebox{90}{URLs}} 
	& plain-HTTP Final URL  
	& 232 / 232 	 & (\fpeval{round((232/232)*100,2)}\%) 
	& 2822 / 153761  & (\fpeval{round((2822/153761)*100,2)}\%) 
	& & 148 / 211 	 & (\fpeval{round((148/211)*100,2)}\%) 
	& 56384 / 152836 & (\fpeval{round((56384/152836)*100,2)}\%) 
	&& 232 / 232     & (\fpeval{round((232/232)*100,2)}\%) 
	& 81405 / 153761 & (\fpeval{round((81405/153761)*100,2)}\%) \\  
	
	& plain-HTTP Inter. URL  
	& 434 / 501 	 & (\fpeval{round((434/501)*100,2)}\%) 
	& 2317 / 149012  & (\fpeval{round((2317/149012)*100,2)}\%) 
	&& 376 / 492 	 & (\fpeval{round((376/492)*100,2)}\%) 
	& 53254 / 147729 & (\fpeval{round((53254/147729)*100,2)}\%) 
	&& 501 / 501     & (\fpeval{round((501/501)*100,2)}\%) 
	& 76656 / 149012 & (\fpeval{round((76656/149012)*100,2)}\%)\\  
	
	& Incompatible Final URL  
	& 280 / 280 	 & (\fpeval{round((280/280)*100,2)}\%) 
	& 2576 / 155689  & (\fpeval{round((2576/155689)*100,2)}\%)
	&& 249 / 263     & (\fpeval{round((249/263)*100,2)}\%) 
	& 55818 / 154501 & (\fpeval{round((55818/154501)*100,2)}\%) 
	&& 280  / 280     & (\fpeval{round((280/280)*100,2)}\%) 
	& 83333 / 155689  & (\fpeval{round((83333/155689)*100,2)}\%) \\
	
	& Incompatible Inter. URL  
	& 193 / 205  	 & (\fpeval{round((193/205)*100,2)}\%)
	& 2724 / 160378  & (\fpeval{round((2724/160378)*100,2)}\%)
	&& 185 / 201     & (\fpeval{round((185/201)*100,2)}\%) 
	& 58082 / 159026 & (\fpeval{round((58082/159026)*100,2)}\%) 
	&& 205  / 205 	 & (\fpeval{round((205/205)*100,2)}\%)
	& 88022 / 160378 & (\fpeval{round((88022/160378)*100,2)}\%) \\
	
	
	\midrule 
	\multirow{3}{*}{\rotatebox{90}{Headers}} 
	& No-HSTS 
	&  200 / 517   	& (\fpeval{round((200/517)*100,2)}\%) 
	& 820  / 31062  & (\fpeval{round((820/31062)*100,2)}\%) 
	&& 462 / 516    & (\fpeval{round((462/516)*100,2)}\%) 
	& 13245 / 30945 & (\fpeval{round((13245/30945)*100,2)}\%) 
	&& 494 / 517    & (\fpeval{round((494/517)*100,2)}\%) 
	& 19842 / 31062 & (\fpeval{round((19842/31062)*100,2)}\%)\\
	
	& No-CSP  
	& 110 / 481 	& (\fpeval{round((110/481)*100,2)}\%) 
	& 463 / 10403 	& (\fpeval{round((463/10403)*100,2)}\%) 
	&& 442 / 481 	& (\fpeval{round((442/481)*100,2)}\%) 
	& 4111 / 10382 	& (\fpeval{round((4111/10382)*100,2)}\%) 
	&& 449 / 481 	& (\fpeval{round((449/481 )*100,2)}\%) 
	& 7246 / 10403 	& (\fpeval{round((7246/10403)*100,2)}\%)\\
		
	& No-Secure in Set-Cookie  
	& 32  / 223  	& (\fpeval{round((32/223)*100,2)}\%) 
	& 236 / 13684 	& (\fpeval{round((236/13684)*100,2)}\%) 
	&& 164 / 222  	& (\fpeval{round((164/222)*100,2)}\%) 
	& 7036 / 13637  & (\fpeval{round((7036/13637)*100,2)}\%)
	&& 142 / 223   	& (\fpeval{round((142/223)*100,2)}\%) 
	& 6979 / 13684  & (\fpeval{round((6979/13684)*100,2)}\%)\\	
	\midrule 
	
	
	\multirow{6}{*}{\rotatebox{90}{TLS}} 
	& Version \textless TLS~1.3   
	& 77 / 579 		& (\fpeval{round((77/579)*100,2)}\%) 
	& 389 / 35399 	&(\fpeval{round((389/35399)*100,2)}\%) 
	&& 547 / 579 	& (\fpeval{round((547/579)*100,2)}\%) 
	& 27867 / 35399 & (\fpeval{round((27867/35399)*100,2)}\%) 
	&& 532 / 579 	& (\fpeval{round((532/579)*100,2)}\%) 
	& 17038 / 35399 & (\fpeval{round((17038/35399)*100,2)}\%) \\
	
	& Version \textless TLS~1.2   
	& 8 / 54 		 & (\fpeval{round((8/54)*100,2)}\%) 
	&  2793 / 149939 & (\fpeval{round((2793/149939)*100,2)}\%) 
	&& 26 / 54 		 & (\fpeval{round((26/54)*100,2)}\%) 
	& 56283 / 149939 & (\fpeval{round((56283/149939)*100,2)}\%) 
	&& \cellcolor{gray!50} 21 / 54 
		& \cellcolor{gray!50} (\fpeval{round((21/54 )*100,2)}\%) 
	& \cellcolor{gray!50} 80153 / 149939 
		& \cellcolor{gray!50} (\fpeval{round((80153/149939)*100,2)}\%) \\
	
	& Non-FS   
	& 13 / 100 		 & (\fpeval{round((13/100)*100,2)}\%) 
	& 2714 / 146180  & (\fpeval{round((2714/146180)*100,2)}\%) 
	&& 43 / 100 	 & (\fpeval{round((43/100)*100,2)}\%) 
	& 55964 / 146180 & (\fpeval{round((55964/146180)*100,2)}\%) 
	&& 57 / 100  	 & (\fpeval{round((57/100)*100,2)}\%) 
	& 77903 / 146180 & (\fpeval{round((77903/146180)*100,2)}\%)\\
	
	& Non-FS+Non-AE   
	& 9 / 71 		 & (\fpeval{round((9/71)*100,2)}\%) 
	& 2615 / 139558  & (\fpeval{round((2615/139558)*100,2)}\%) 
	&& \cellcolor{gray!50} 25 / 71 
		& \cellcolor{gray!50} (\fpeval{round((25/71)*100,2)}\%) 
	& \cellcolor{gray!50} 55662 / 139558 
		& \cellcolor{gray!50} (\fpeval{round((55662/139558)*100,2)}\%) 
	&& 38 / 71 		 & (\fpeval{round((38/71)*100,2)}\%)
	& 74241 / 139558 & (\fpeval{round((74241/139558)*100,2)}\%)\\
	
	& Expired Cert.    
	& 3 / 21   		 & (\fpeval{round((3/21)*100,2)}\%)
	& 2804 / 149945  & (\fpeval{round((2804/149945)*100,2)}\%) 
	&& 15 / 21 		 & (\fpeval{round((15/21)*100,2)}\%) 
	& 56325 / 149945 & (\fpeval{round((56325/149945)*100,2)}\%) 
	&& \cellcolor{gray!50} 6 / 21 
		& \cellcolor{gray!50} (\fpeval{round((6/21)*100,2)}\%) 
	& \cellcolor{gray!50} 80499 / 149945 
		& \cellcolor{gray!50} (\fpeval{round((80499/149945)*100,2)}\%)\\
	
	& Invalid Cert. Host Name  
	& 31/54 		 & (\fpeval{round((31/54)*100,2)}\%) 
	&  2767  / 142126 & (\fpeval{round((2767/142126)*100,2)}\%) 
	&& 39	/	54    & (\fpeval{round((39/54)*100,2)}\%) 
	& 55703 / 142126  & (\fpeval{round((55703/142126)*100,2)}\%) 
	&& 37/54  	 	& (\fpeval{round((37/54)*100,2)}\%) 
	& 79950 / 142126 & (\fpeval{round((79950/142126)*100,2)}\%)\\
	
	\bottomrule
\end{tabular}
\end{adjustbox}
\end{table*}

\section{Attack Scenarios} \label{sec:scenario}
We now provide two attack scenarios that can benefit from the regional HTTPS security inconsistencies.

\subsection{Region-Confusion Attack} 
\begin{figure}[tp!]
	\centering
		\includegraphics[width=\linewidth]{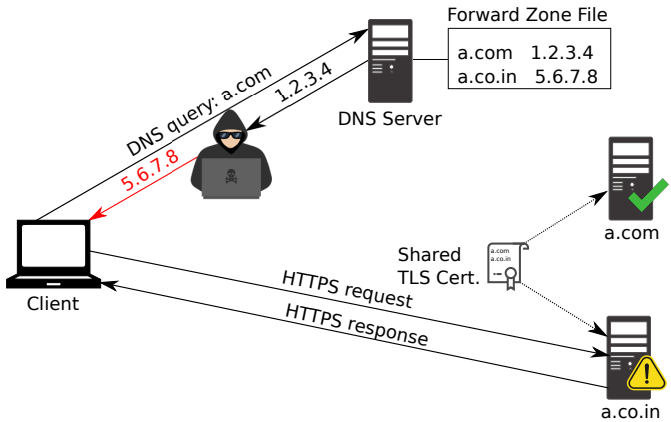}
	\caption{The region-confusion attacker model.}
	\label{fig:region}
\end{figure}

In this attack, the attacker is located on the communication channel and has control over it. It can passively eavesdrop, or actively inject, modify, drop, replay, or redirect messages, sent from, or to, the client or server. We assume that the domain has regional HTTPS security inconsistencies. That is, if two clients at different regions request the same domain (e.g. \texttt{example.com}), some regions receive weak security guarantees, while other regions receive strong guarantees. The requested domain may perform URL redirection based on the region (\quotes{regional redirection}) to a different domain that may have a different IP, contains a new or different subdomain, different TLD, or different documents. We assume the redirection occurs form gTLDs to ccTLDs, and not between ccTLDs. For example, requesting \quotes{example.com.sg} from the UK does not redirect to any other domain, while requesting \quotes{example.com} redirects to \quotes{example.co.uk}. We also assume that the requested domain uses the same user credentials (e.g. username and password) such that users can login to multiple domains with the same credentials. Our attacker aims to force the user, e.g. via DNS spoofing if the targeted domains share the certificate, to connect to the weaker domain that has weaker security guarantees. For example, the attacker redirects a UK-based client's request for \texttt{example.com} from going to \texttt{example.co.uk} to another regional domain that belongs to the same domain, but provides weaker HTTPS security guarantees, e.g. \texttt{example.co.br}. This allows the attacker to exploit the weaknesses that exist in the weak region's domain, e.g. perform an XSS attack, due to the absence of a security configuration, e.g. the CSP header, or perform TLS stripping attack, due to the absence of the HSTS header combined with the absence of the includeSubDomains (if the HSTS is present in the parent domain and the regional domain is a subdomain of the parent domain). In the same vein, the absence of the FS property can enable long-term mass surveillance since the attacker can decrypt past session keys whenever the long-term non-FS key is broken. The absence of the AE property can enable some attacks over the MAC-then-Encrypt schemes as shown in~\cite{alfardan13,vaudenay02}. Moreover, using legacy versions of the TLS protocol can enable some known attacks that has been batched in the newer versions including downgrade attacks~\cite{alashwali18}. Finally, expired certificates and invalid hostnames can be abused by man-in-the-middle attackers that can confuse users with forged certificates, thus, they should not be used by legitimate website. This attacker model allows more persistent, and harder to detect attacks than in the classical phishing attacks, e.g. via a faked website. This is because, in this attack, the user is redirected to a legitimate website that accepts the user's credentials. The attacker can obtain and abuse, in the long-term, the user's credentials, as opposed to a faked phishing website, where the user is normally more likely to find out that the website is faked, due to the absence of real content. \autoref{fig:region} illustrates the region-confusion attacker model.


\subsection{Discrimination Attack} 
\begin{figure}[tp!]
	\centering
	\includegraphics[width=\linewidth]{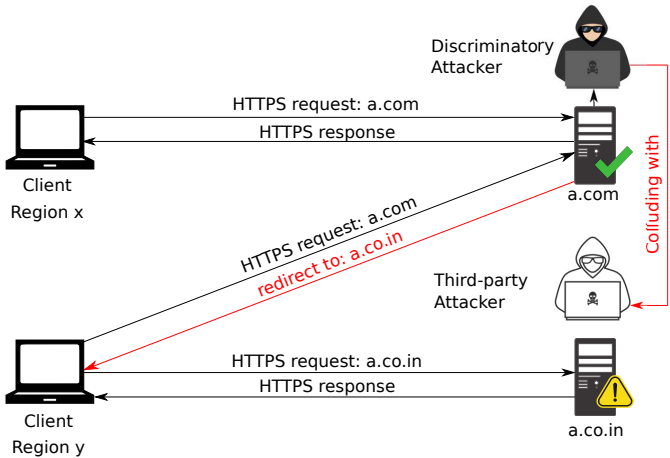}
	\caption{The discriminatory attacker model.}
	\label{fig:discriminatory}
\end{figure}

This attack scenario is based on the \quotes{discriminatory} adversarial model, which is initially introduced in~\cite{alashwali19fs}. This attacker is located on the server side. It can be represented by an insider attacker (dishonest system administrator), a dishonest organisation (e.g. giant cloud provider), or a malware that hit the server. It weakens the security guarantees provided to clients in some regions, for a powerful third-party's advantage, e.g. a state-level attacker, who has the capabilities to exploit this weakness. For example, the discriminatory attacker can deny some regions from some security guarantees such as the FS property, to enable the powerful third-party attacker to decrypt collected traffic either at present, due to the third party's powerful capabilities of breaking non-FS keys, or in the future once the server's long-term key is broken due to advancement in computing power, or if the key is given to the powerful third-party attacker, after the key is expired when it is no longer used by the server. The semi-trusted server colludes with, or compelled by, a powerful third-party attacker. This model gives the server a financial, legal, and reputation advantage over giving every session key or the decrypted traffic to the third-party powerful attacker. This attacker model is inspired by real-world incidents such as the \quotes{export-grade} cryptography, a depreciated US law that mandated weaker cryptography to products, including software, exported outside the US \cite{export18}, in addition to the alleged \quotes{PRISM} program in which giant service providers open back-doors that are known to third parties (e.g. intelligence) \cite{prism18}. \autoref{fig:discriminatory} illustrates the discriminatory attacker model.


\section{Discussion} \label{sec:discussion}
Based on our analysis and observations in this experiment, we provide the following discussion:
\begin{enumerate}
\item \textbf{Security favours simplicity:}
as shown in \autoref{sec:relationship}, domains with inconsistent HTTPS security tend to have higher percentages of URLs and IPs diversity, and to a lesser extent higher redirections, compared to domains with secure HTTPS. This suggests that the more complex the domain is (e.g. diverse IPs and URLs) the more inconsistent HTTPS security is.

\item \textbf{The need for secure redirection, and avoiding redirection if possible:}
the security of intermediate URLs in redirections is no less important than the security of final URLs. Therefore, all intermediate connections need to be over TLS and need to deploy the same security measures as final URLs. This is also noted in~\cite{chang17}, where they suggest strict redirection policy in the same vein as HSTS. We also recommend users to use the regional domain directly in their requests if known, as this can provide better security than requesting the generic domain that redirects to the regional domain, to avoid insecure redirections, as shown in the case of \texttt{match.com} in \autoref{sec:urls}. Additionally, avoiding redirections can be automated by a browser extension that buffers the URL chain and makes the request to the final URL, and by prioritizing regional domains in search engines. 

\item \textbf{TLS is not only about secrecy:}
while blocking pages content is public and may not require secrecy, the authenticity and integrity of such pages are still important to avoid Denial of Service (DoS) attacks for example. Therefore, domain owners and governments need to be aware of the TLS goals and the possible harm that can result from unauthenticated blocking pages and the users' habituation to them.

\item \textbf{HTTPS security inconsistencies and the potential effect on Tor network users:}
the Onion Router (Tor) network aims to provide anonymity to its users. It hides users' locations and the websites they visit online, by routing traffic through multiple relay servers run by volunteers all over the world~\cite{tor19}. In Tor, if the domain names are resolved using \quotes{SOCKS 4a}, which passes hostnames to relays, or through \quotes{Tor-resolve}, which uses Tor network to resolve host names remotely~\cite{tor19}, this means that Tor users may be redirected to regional domain names that are different from their region. As our results show, there is evidence of HTTPS security inconsistencies in servers' responses to requests from different regions. Therefore, a user may end up connecting to a weaker version of the intended website, mainly because the generic domains, e.g. \texttt{example.com} are resolved in another region, e.g. \texttt{in.example.com}, which may have weaker HTTPS security.

\item \textbf{Regional domain format:}
a single standard format for regional domains, with the same-origin policy in mind, can help in reducing phishing attacks' surface and help users identify malicious domains and redirections. The domain same-origin policy can be achieved in brand TLDs, where brands are used as a TLD instead of the traditional TLDs, and in regional subdomains, but not in ccTLDs.

\item \textbf{Tools for consistency and redirection testing:}
tools for consistency assessment may help administrators and users in mitigating and detecting HTTPS security inconsistencies. 
\end{enumerate}
\section{Related Work} \label{sec:related}
In recent years, several studies have looked at servers' responses inconsistencies among clients that differ in some aspects such as client type, vantage point, or geographic location. However, to the best of our knowledge, our study is the first that reports on HTTPS security inconsistencies among different regions. In what follows, we summarise relevant work. \par 

In terms of inconsistencies from the client type aspect, Mendoza et al. \cite{mendoza18} analysed the inconsistencies of security headers in servers' responses to mobile versus desktop users. Out of the \num{70000} examined Alexa's top domains, they found overall \num{2000} domains with inconsistencies in one or more of the examined security headers. Khattak et al.~\cite{khattak16} measured the inconsistencies of servers' responses to anonymous (e.g. using Tor) versus normal users. They report 1.3  million addresses of the  IPv4 address space and around 3.67\% of the top \num{1000} Alexa domains either block or provide a degraded service to Tor users. \par  

In terms of inconsistencies from the regional aspect, Afroz et al. \cite{afroz18} studied server-side blocking of regions. They confirmed the existence of servers that block users based on region. Several recent studies such as Samarasinghe and Mannan~\cite{samarasinghe19}, and Fruchter et al.~\cite{fruchter15} reported inconsistencies in users' privacy such as the use of third party trackers between clients from different regions. Eijk et al.~\cite{eijk19} studied a similar aspect and found that the inconsistencies in cookie notices are related to TLDs and not the user location. However,~\cite{eijk19} pointed out this conclusion does not hold with \texttt{com} domains, where the location relates to cookie notices inconsistencies, because these generic TLDs perform redirection based on the location. Niaki et al.~\cite{niaki19} provide a framework for measuring Internet censorship against users from different regions. \par 

In terms of inconsistencies from vantage point aspect, Jueckstock et al.~\cite{jueckstock19} examined the consistency of servers' responses to requests initiated from different vantage points (VP): cloud data centre, research university, residential network, and Tor gateway proxy. They found slight differences, and they recommend university VP over cloud providers VP in measurement studies, as the university VP generalises \quotes{slightly better} to the actual residential browsing experience. \par 

From TLS configurations and HTTPS redirections aspect, Amann et al.~\cite{amann17} conducted TLS scans from various locations. They noticed a small fraction of inconsistencies in the HSTS and HPKP headers in the active scans, but they do not analyse them. Unlike Amann et al. who do not follow redirections and connect to the \quotes{base domain} in all regions, we do follow HTTPS redirections, then, we make TLS handshakes with the final response domains at each region (which can differ between regions). Following redirections is more representative of web browsers behaviour, and revealed inconsistencies. In terms of HTTPS redirections security, Chang et al. \cite{chang17} examined Alexa's top 1 million domains and found that the majority (83.3\%) of HTTPS redirections in TLS servers are insecure. Alashwali et al.~\cite{alashwali19} measured the differences in TLS configurations and certificates between plain-domains and their equivalent www-domains. They found that www-domains tend to be more secure than plain-domains. In addition, they found that over 50\% of the HTTPS plain-domains that showed weaker configurations than their equivalent www-domain are redirected to www-domains. However, these redirections are not always secure.

\section{Limitations and Future Work} \label{sec:limitations}
In terms of limitations, first, due to the dynamic and rapidly changing nature of the web, the domains and examples included in this study may have changed their configurations over time. This is a well-understood limitation in Internet measurement studies in general. However, even if some of those websites have changed, this does not diminish the value of the insights we gained from analysing those data. Additionally, this does not eliminate the existence of inconsistencies in another set of websites. Second, in order to scale, the HTTP and TLS connections to the domains were automated through programmable HTTP and TLS clients. These tools simulate an HTTPS client such as a web browser. However, we do not advertise a specific browser vendor in the requests' headers. Nevertheless, the domains we present in the tables or figures in this paper are cross-validated against the latest version of the Chrome browser at the time of the study. Additionally, we measure the consistency of servers' responses between different regions despite the client's vendor. We use the same client in all regions. Thus, our results should not be affected if a server provides different responses to different vendors. Third, in headers and TLS security inconsistencies, we check the values provided in the final response only and we do not check the intermediate URLs. Analysing each URL in the redirection chain is time and resource prohibitive for this project. Fourth, we analyse security headers in terms of headers' presence/absence only. Checking the headers' values such as syntax and configurations correctness is a further level in the analysis depth that is outside our paper's scope, and we leave it to future work. Fifth, we do not analyse the responses page content except the manual analysis we did in \autoref{sec:urls} to identify blocking home pages. Finally, we check our dataset domains (requests' domains) against known malicious domain by Google's safe browsing, and we exclude the responses of malicious domains requests from our analysis. However, we do not check intermediate or final response domains against malicious domains nor exclude them, if any. Future work can look at malicious final and intermediate response URLs, and their inconsistencies, possibly using new methods for detecting malicious URLs such as the one introduced in~\cite{kim18}.

\section{Conclusion} \label{sec:conclusion}
In this paper, we demonstrated the existence of a previously unexplored phenomenon. That is, the existence of HTTPS security inconsistencies in servers' responses to clients located in five different regions. We quantified the HTTPS security inconsistencies we identified. These inconsistencies can provide a false sense of security among users from different locations, and can enable attacks that redirect the user's request to the weaker region to exploit the weaker region's weaknesses. We draw the recommendations from our experiment observations, which suggest standardising regional domain format and secure redirection, in addition to the need for testing tools for domain administrators and users that help to mitigate and detect regional domains' inconsistencies.


\bibliographystyle{IEEEtran}
\bibliography{refs}
\end{document}